\documentclass[proof]{WileyASNA-v1}

\articletype{Article Type}

\def \src {\mbox{A0538$-$66}}
\def \xmm {\textit{XMM-Newton}}
\def \lum {erg\,s$^{-1}$}

\def \cou {cts\,s$^{-1}$}
\def \brem {bremsstrahlung}

\newcommand{\pdx}[1]{_{\rm #1}}

\received{30 August 2024}
\revised{30 August 2024}
\accepted{30 August 2024}

\begin{document}

\title{\xmm\ observations of the peculiar Be X-ray binary \src}

\author[1,2]{Michela Rigoselli*}

\author[2,3]{Caterina Tresoldi}

\author[4,5]{Lorenzo Ducci}

\author[2]{Sandro Mereghetti}

\authormark{RIGOSELLI \textsc{et al}}

\address[1]{\orgdiv{INAF}, \orgname{Osservatorio Astronomico di Brera}, \orgaddress{\state{via Brera 28, 20121 Milano}, \country{Italy}}}

\address[2]{\orgdiv{INAF}, \orgname{Istituto di Astrofisica Spaziale e Fisica Cosmica di Milano}, \orgaddress{\state{via A. Corti 12, 20133 Milano}, \country{Italy}}}

\address[3]{\orgdiv{Dipartimento di Fisica}, \orgname{Università degli Studi di Milano},
\orgaddress{\state{Via Celoria 16, 20133 Milano}, \country{Italy}}}

\address[4]{\orgdiv{Institut für Astronomie und Astrophysik, Kepler Center for Astro and Particle Physics}, \orgname{Universität Tübingen}, \orgaddress{\state{Sand 1, 72076 Tübingen}, \country{Germany}}}

\address[5]{\orgdiv{ISDC Data Center for Astrophysics}, \orgname{Université de Genève}, \orgaddress{\state{16 chemin d’Écogia, 1290 Versoix}, \country{Switzerland}}}

\corres{* \email{michela.rigoselli@inaf.it}}


\abstract{\src\ is a neutron star/Be X-ray binary located in the Large Magellanic Cloud and, since its discovery in the seventies, it showed a peculiar behavior which makes it a unique object in the high-mass X-ray binaries scene: the extremely eccentric orbit ($e=0.72$), the short spin period of the neutron star ($P=69$ ms), the episodes of super-Eddington accretion. These characteristics contribute to a remarkable bursting activity that lasts from minutes to hours and increases the flux by a factor $10^3-10^4$.
In 2018, \src\ was observed by \xmm\ in a particularly active state, characterized by a forest of short bursts lasting $0.7-50$ seconds each.
In this contribution we present a reanalysis of these observations. The timing analysis allowed us to distinguish between the epochs of direct accretion and propeller state, that do not correlate with the orbital position of the neutron star. The spectral analysis revealed that during the accretion regime three components (a soft one, a hard one, and a $\sim\!6.4$-keV emission line) equally contribute to the overall emission, while the propeller regime is characterized by a single soft component.
We discuss these findings in the context of spherical and disk accretion regimes, highlighting the similarities and the differences with other X-ray binary systems.}



\fundingInfo{Funding info text.}

\maketitle

\section{Introduction}\label{sec:intro}
\src\ is a neutron star/Be X-ray binary (BeXB) located in the Large Magellanic Cloud (LMC). 
The neutron star (NS) was detected pulsating only once, in 1980, when it was in super-Eddington accretion regime. The measured spin period was $P_s=69$ ms with a pulsed fraction of about $10\%$ \citep{1982Natur.297..568S}.

\begin{figure*}[t]
\centerline{\includegraphics[height=8cm]{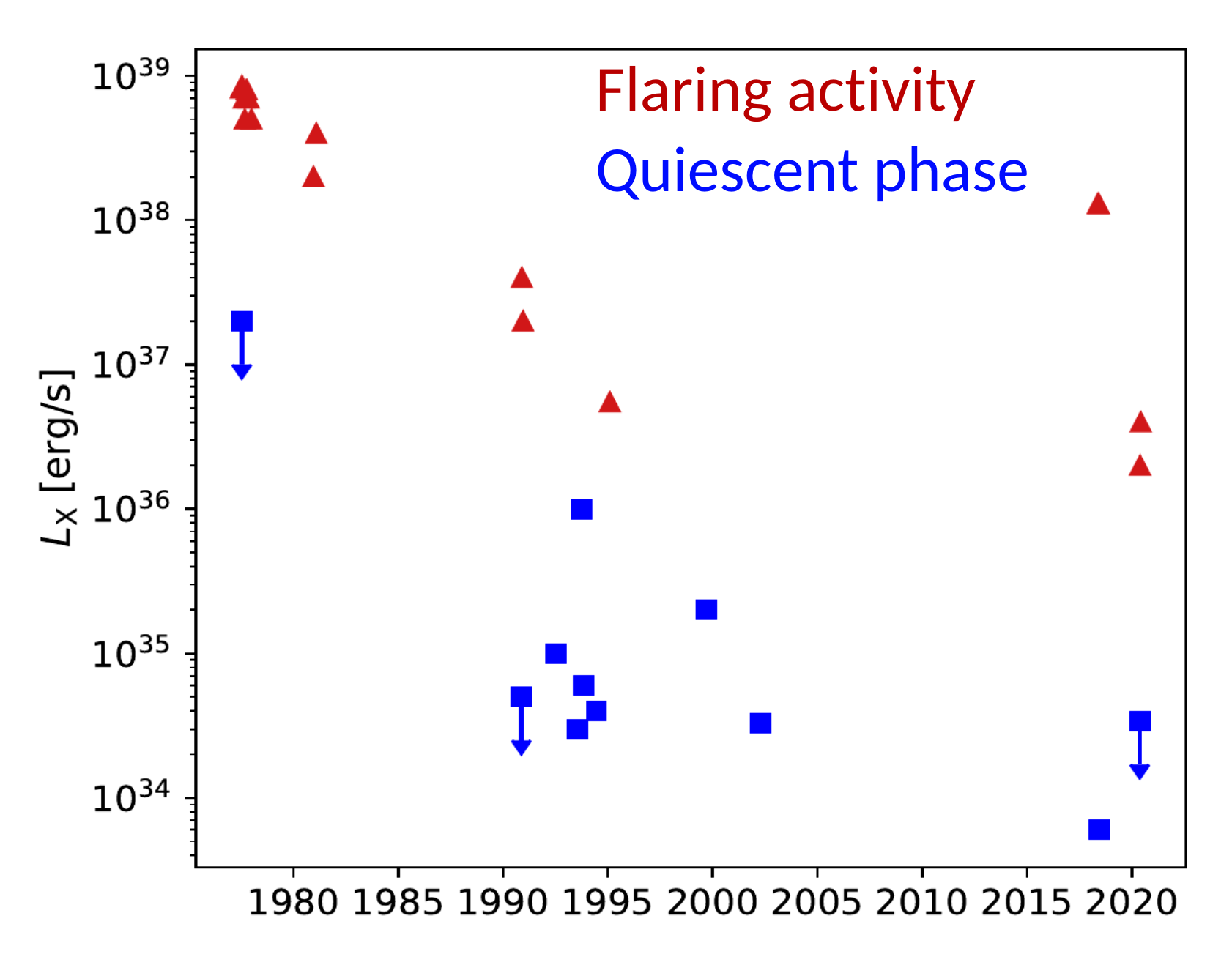} \qquad
\includegraphics[height=8cm]{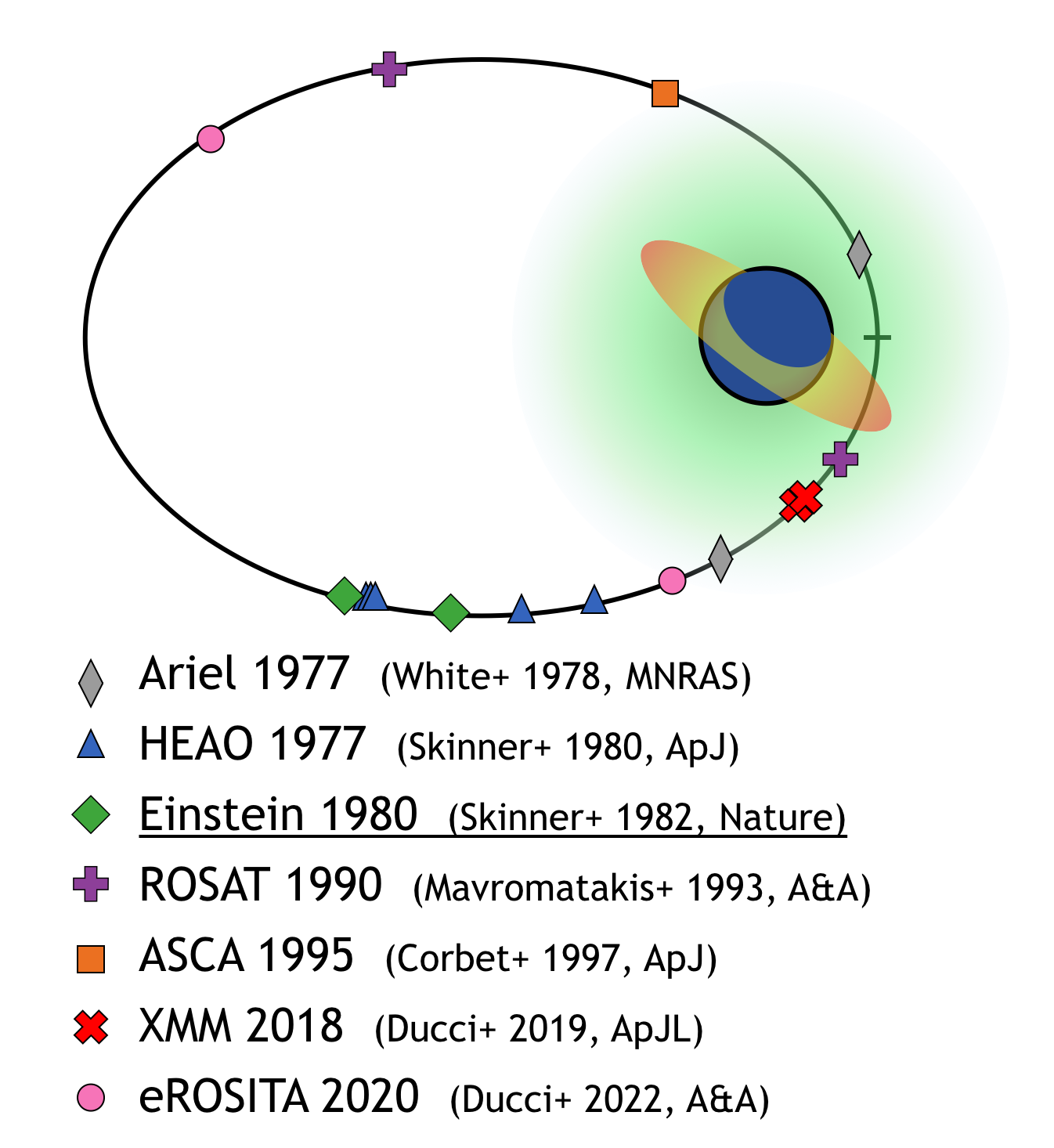}}
\caption{Left panel: \src\ X-ray luminosity attained in various observations since its discovery; the red triangles represent the luminosity reached during the flaring activity, while the blue squares correspond to the quiescent brightness level, which is constant throughout the observation. 
Right panel: positions of the NS in the orbit during observations in which flaring activity was recorded.\label{fig:orbit}}
\end{figure*}

Since its discovery in the late Seventies \citep{1978MNRAS.183P..11W,1980ApJ...240..619S}, \src\ was observed many times with various X-ray satellites, and it was detected in two main states: \textit{(i)} flaring activity phase, having a luminosity in the range $10^{36}-10^{39}$ \lum\ and lasting from hours to days (Figure~\ref{fig:orbit}, left panel, red triangles); \textit{(ii)} quiescent phase, characterized by a dimmer ($10^{34}-10^{36}$ \lum), constant emission throughout the whole observation (Figure~\ref{fig:orbit}, left panel, blue squares).

\begin{figure}[t]
\centerline{\includegraphics[width=1\columnwidth]{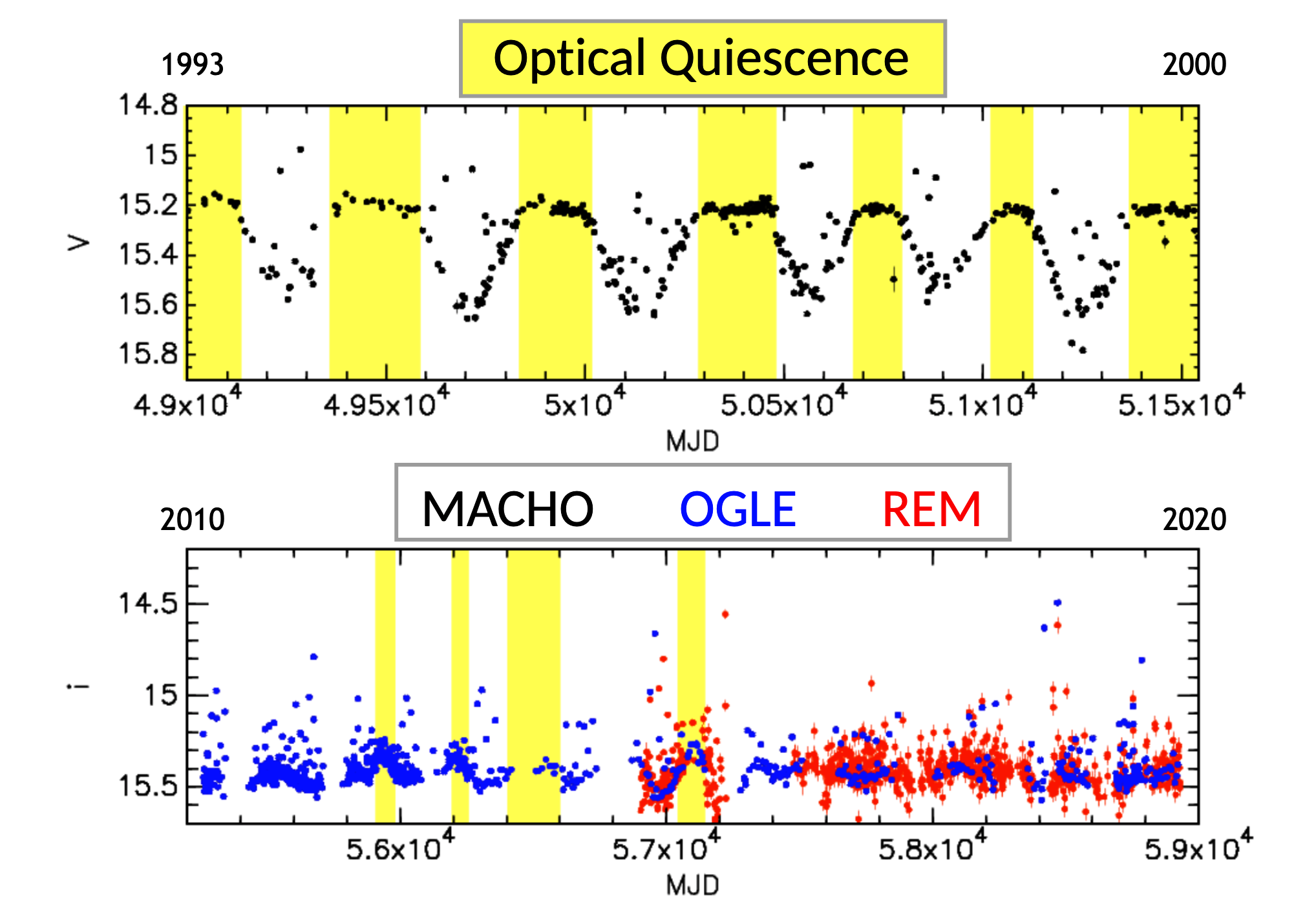}}
\caption{MACHO, OGLE, and REM optical light curves of \src. Yellow regions show the quiescent optical states, when the flaring activity is absent. Adapted from \citet{2022A&A...661A..22D}.\label{fig:optical}}
\end{figure}

The X-ray activity of the source does not correlate with the orbital phase, as the right panel of Figure~\ref{fig:orbit} illustrates, but with the presence of the circumstellar decretion disk surrounding the Be companion star. As Figure~\ref{fig:optical} shows, the optical light curve reveals an alternation between epochs of bright and constant optical emission (called “quiescent optical state” by \citealt{2001MNRAS.321..678A,2003MNRAS.339..748M}), and epochs with a dimmer but more variable optical emission, associated to the occultation of the Be star by its circumstellar disk.
When the neutron star crosses the circumstellar disk, where the outflow from the Be star is denser and slower ($\dot{M} \sim 100-1000$ higher and $v_\infty \sim 10-100$ lower than in the polar wind region; e.g. \citealt{1988A&A...198..200W}), the higher accretion rate produces X-ray flares, and consequently, optical flares from reprocessing of X-ray photons \citep{2019A&A...624A...9D}.
The alternation between these two states has been  regular and quasi-periodic until 2000, but more recently the quiescent optical state seems to be vanished. This  might be explained by the presence of a circumstellar disk that has become more stable over the years.

\xmm\ observed \src\ four times: once in 2002, when the NS was far from  periastron (ObsID 007174, referred to as obs.\ 0), and in three consecutive orbits in 2018, when the NS was close to  periastron  (ObsID 082348, obs.\ A, B and C). 
Previous works \citep{2004ESASP.552..329K,2019ApJ...881L..17D} found a peculiar bursting activity in obs.\ A and B, and constant emission throughout obs.\ 0 and C. \citet{2019ApJ...881L..17D} interpreted the onset of rapid bursts in obs.\ A and B with spherically symmetric accretion modulated by a magnetic gating mechanism.

Here we present a reanalysis of the four \xmm\ observations (Section~\ref{sec:data}), studying the flares with the Bayesian block segmentation method (Section~\ref{sec:timing}), and performing a count-rate resolved spectral analysis (Section~\ref{sec:spec}). In Section~\ref{sec:disc} we discuss our findings and the emission regimes of the source. Conclusions are set in Section ~\ref{sec:conc}.

\section{Observations and data reduction}\label{sec:data}

\setlength{\tabcolsep}{0.75em}
\begin{table*}
\centering \caption{Journal of the \xmm\ observations of \src}
\label{tab:log}

\footnotesize
\begin{tabular}{llcccccccc}
\toprule
\midrule
Nick & Obs. ID & Start time & End time  & Orbital phase & \multicolumn{3}{c}{Net Exposure time (ks)} \\[3pt]
         & & (UTC)        & (UTC)      &  & EPIC-pn & EPIC-MOS1 & EPIC-MOS2  \\[3pt]
\midrule
0 & 0071740501 & 2002-04-09 21:01:47 & 2002-04-10 02:53:36 & 0.637--0.651 & 11.2 &    - & 17.2 \\
A & 0823480201 & 2018-05-15 06:31:09 & 2018-05-15 11:11:50 & 0.958--0.971 &  5.1 &  8.1 &  8.4 \\
B & 0823480301 & 2018-05-31 22:31:03 & 2018-06-01 02:05:03 &  0.959--0.970 & 4.5 &  7.2 &  7.2 \\
C & 0823480401 & 2018-06-17 13:00:41 & 2018-06-17 16:29:42 &  0.957--0.967 & 8.5 & 12.3 & 12.1 \\
\bottomrule\\[-5pt]
\end{tabular}

\raggedright
\end{table*}

In the four observations of \src\ (see Table \ref{tab:log}) the EPIC cameras were operated in Small Window mode (time resolution 0.3 s for EPIC-MOS, 5.7 ms for EPIC-pn) with thin optical filter. 

The data reduction was performed using the \texttt{epproc} and \texttt{emproc} pipelines of version 20 of the Science Analysis System (SAS)\footnote{https://www.cosmos.esa.int/web/xmm-newton/sas}. 
Time intervals of high background were removed excluding periods with count rate above $0.2$ \cou\ in the energy range $10$--$12$ keV. The resulting net exposure times are given in Table~\ref{tab:log}.

We selected single- and multiple-pixel events (\textsc{pattern}$\leq$4 for the EPIC-pn and $\leq$12 for the -MOS); source counts were extracted in a circular region of radius $30''$ centered at R.A\,=\,$05$:$35$:$41.3$ and Dec\,=\,$-66$:$51$:$51$ (J2000), while the background was chosen in circular regions far away from the source.

\section{Results}\label{sec:res}

\subsection{Timing analysis}\label{sec:timing}

\begin{figure}[htb]
\centerline{\includegraphics[width=1\columnwidth]{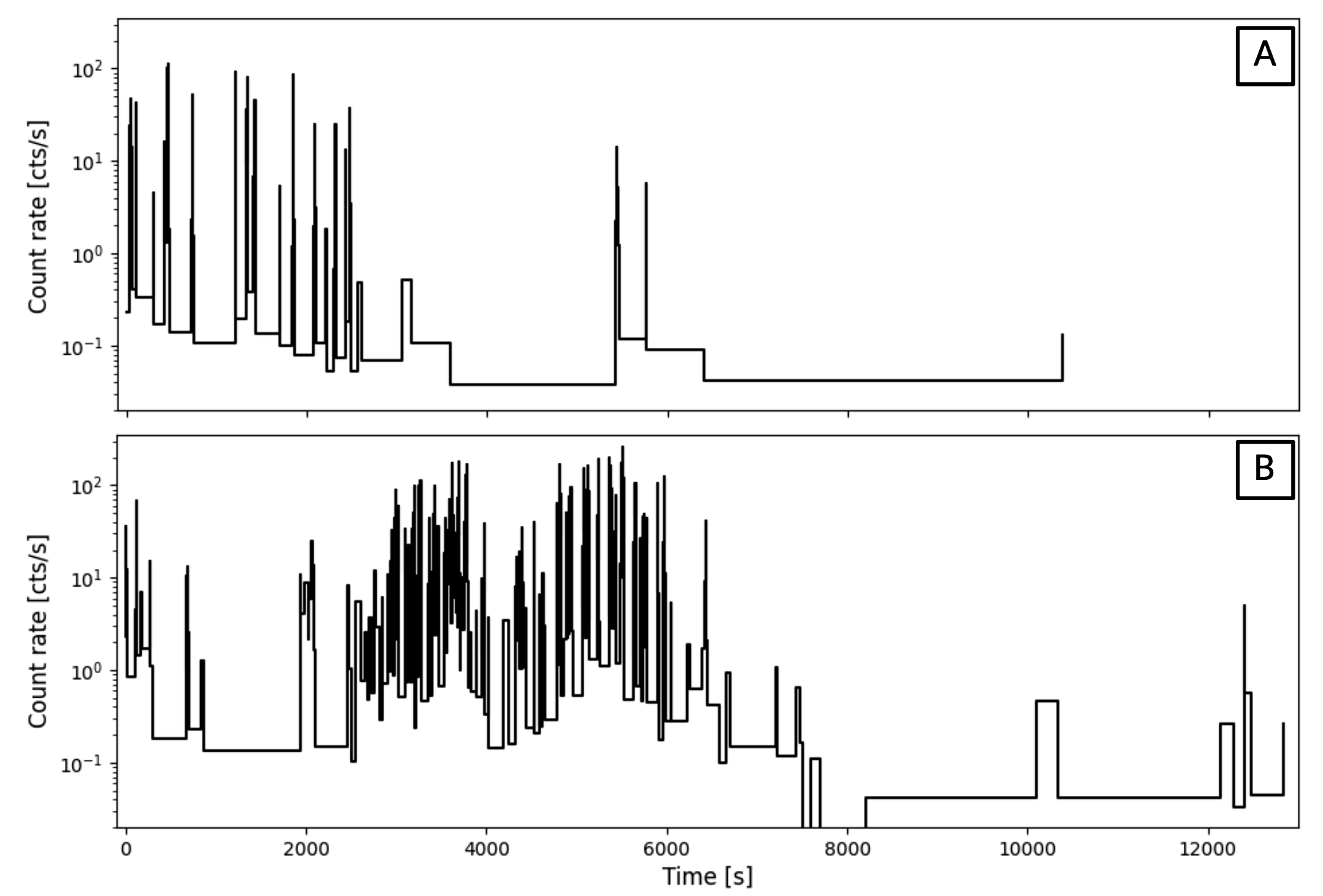}}
\caption{EPIC-pn light curves between $0.3$--$10$ keV of obs.\ A (top panel) and B (bottom panel) obtained by the Bayesian block segmentation method.\label{fig:lc}}
\end{figure}

Figure~\ref{fig:lc} shows the $0.3$--$10$ keV light curves of the EPIC-pn obs.\ A and B obtained by the Bayesian block segmentation method \citep{2013ApJ...764..167S} making use of the \texttt{bayesian\_blocks} task of the Astropy library \citep{2018AJ....156..123A}.
They clearly exhibit a remarkable bursting activity for most of the exposure time. The bursts have different duration from $0.7$ to $50$ seconds, and show an increase of the count rate by a factor $100$--$1000$. At the end of the bursting forest, a constant level of emission of about $0.04$ \cou\ is reached in both observations.

Obs.\ 0 and C, on the contrary, exhibit a constant emission throughout the whole observations and have a count rate in the $0.3$--$10$ keV energy range of $0.09$ \cou\ and $0.03$ \cou, respectively.

\subsection{Spectral analysis}\label{sec:spec}

We used the results of the timing analysis to investigate the spectral variability of the bursting state. To this aim, we extracted four source spectra at different intervals of EPIC-pn count rates (CR) in the following ranges: CR$>$50 \cou\ (high), 25$<$CR$<$50 \cou\ (medium), 5$<$CR$<$25 \cou\ (low), CR$<$0.04 \cou\ (infra-bursts). 

We took into account pile-up effects, that are expected when the source count rate exceeds $>$25~\cou\ in Small Window mode.
For the EPIC-pn, we generated a response file that includes pile-up corrections\footnote{We followed the procedure described in the SAS thread: \url{https://www.cosmos.esa.int/web/xmm-newton/sas-thread-epatplot}.} suitable for each CR level.
Since a response file including pile-up corrections cannot be produced for the EPIC-MOS, during the intervals of high CR we used only the EPIC-pn spectra.

The spectral analysis was performed using XSPEC (version 12.13.0c) and PyXspec (version 2.1.2). The spectra from the EPIC-pn camera were rebinned using the \texttt{specgroup} tool with a minimum of 50 counts per bin.
The spectra of the infra-bursts time intervals of obs.\ A and B, and the whole obs.\ 0 and C were extracted with a maximum likelihood technique, which is more suitable for dim sources \citep[see e.g.][]{2018A&A...615A..73R}. 
After checking that the spectra of the A and B observations were consistent at each count rate level, we merged them with the tool \texttt{epicspeccombine}. 

Given the secure association with the LMC, \src\ has a well known distance of 50 kpc \citep{2004NewAR..48..659A} and a rather low absorption column ($N_{\rm H}\simeq9\times10^{20}$ cm$^{-2}$ along the whole line of sight, \citealt{2016A&A...594A.116H}).
The interstellar medium absorption was fitted with the \textsc{tbvarabs} model, with abundances set to LMC values (see \citealt{2019ApJ...881L..17D} for  details).

In the following, we give the errors at $1\sigma$ confidence level.

\subsubsection{CR resolved bursts}

\begin{figure}[htb]
\centerline{\includegraphics[width=1\columnwidth]{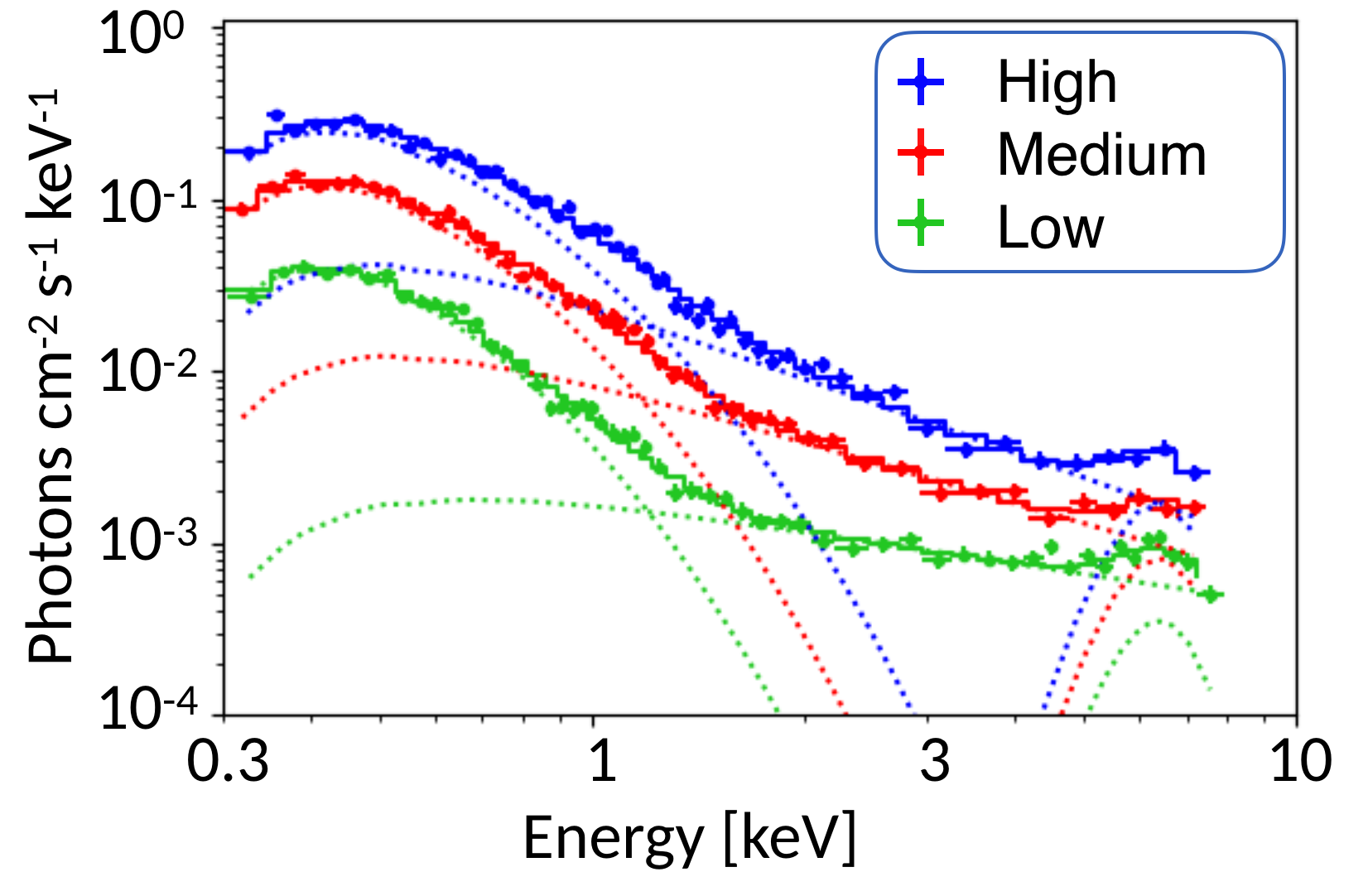}}
\caption{EPIC-pn spectra of the bursts at different CR levels (CR$>$50 \cou: blue; 25$<$CR$<$50 \cou: red; 5$<$CR$<$25 \cou: green) obtained by combining the spectra of obs.\ A and B.  These spectra are fitted with a model (dotted lines) composed by a \brem\ plus a power law and a Gaussian emission line.\label{fig:spec_burst}}
\end{figure}

The three cumulative spectra of the bursts at different CR levels are shown in Figure~\ref{fig:spec_burst}. 
They cannot be fitted by a single-component model. At least two components are needed to fit the continuum, with the addition of a broad emission line (width $\sigma\sim0.7$ keV) at $E=6.4$ keV.

Our best fit\footnote{We included a systematic error of 5\% to account for unmodeled residuals below $\sim\!1.5$ keV.} ($\chi^2$/dof = $172.50/144$) is a model composed by a \brem, that accounts for the softer part of the spectrum, plus a power law at harder energies. We also tried a blackbody plus a power law ($\chi^2$/dof = $176.93/144$) and a disk-blackbody \footnote{The spectrum from an accretion disk consisting of multiple blackbody components, see \citep{1984PASJ...36..741M,1986ApJ...308..635M}.} plus a power law ($\chi^2$/dof = $175.35/144$).

Regardless of the model employed to fit the soft, thermal component, we found that it accounts for $\sim\!30$--$40$\% of the total flux between $0.3$--$10$ keV.
Both the soft and the hard components vary in  shape  as a function of CR. In particular, the temperature increases as the CR increases ($kT$ changes from $0.29\pm0.01$ keV to $0.35\pm0.02$ keV to $0.43\pm0.03$ keV in the \brem\ model), and the photon index steepens ($\Gamma$ changes from $0.60\pm0.07$ to $1.0\pm0.1$ to $1.2\pm0.1$).

The absorption column was fixed to a common value among the three spectra, and it was found to be $N_{\rm H}=(8.0\pm0.7)\times 10^{20}$ cm$^{-2}$. The emission line did not shown any substantial variation between the three CR states, aside from its normalization. We fixed the central energy at $6.4$ keV, and we found a common width of $0.73_{-0.15}^{+0.24}$ keV.

The total luminosities in the $0.3$--$10$ keV energy range are: $(2.45\pm0.05)\times10^{37}$ \lum\ (low CR), $(6.13\pm0.02)\times10^{37}$ \lum\ (medium CR), and $(1.33\pm0.02)\times10^{38}$ \lum\ (high CR).

\subsubsection{Infra bursts}

\begin{figure}[htb]
\centerline{\includegraphics[width=1\columnwidth]{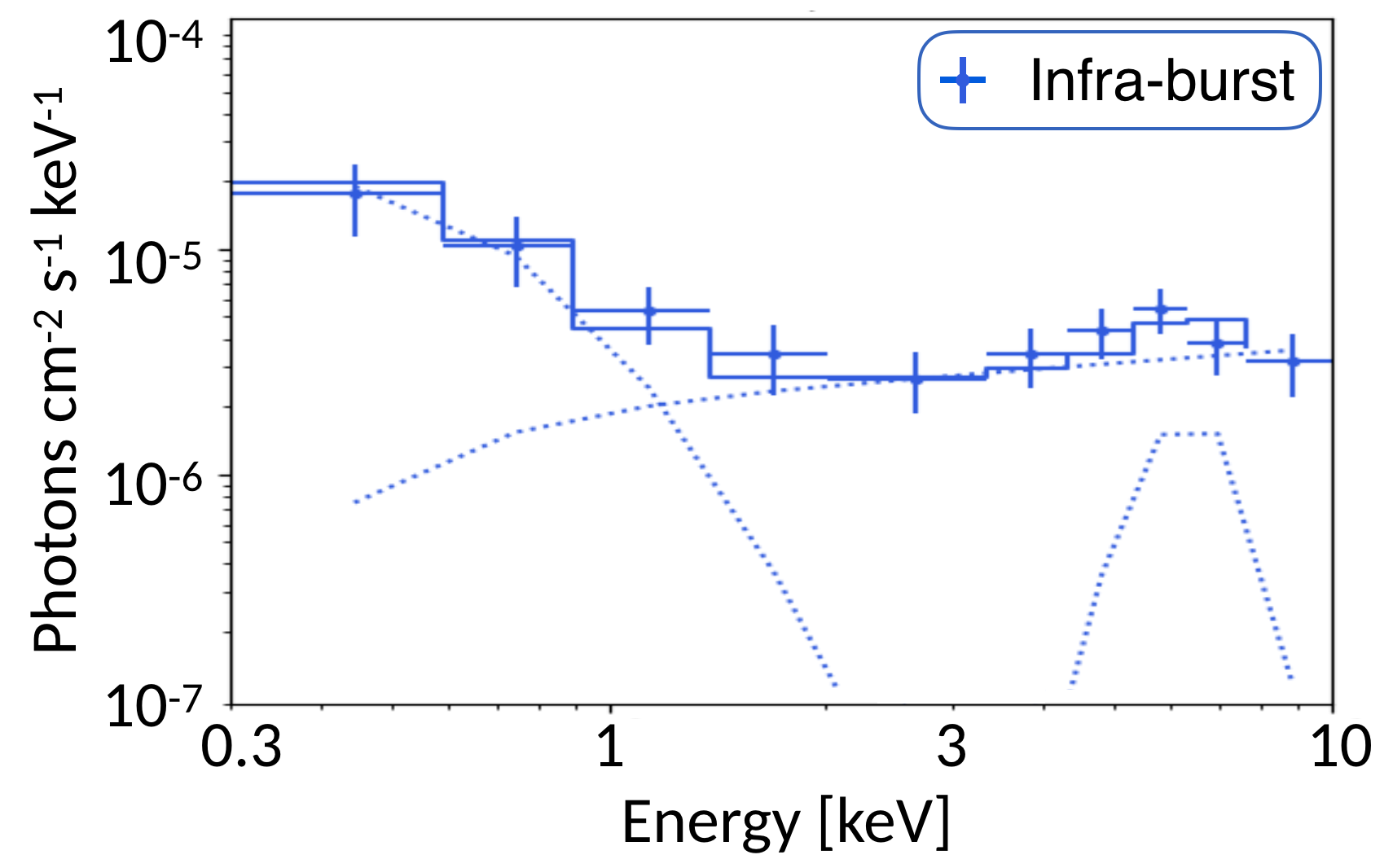}}
\caption{EPIC-pn spectrum of the infra-burst time intervals obtained by combining the spectra of obs.\ A and B. The spectrum is fitted with a model (dotted lines) composed by a \brem\ plus a power law and a Gaussian emission line.\label{fig:spec_infra}}
\end{figure}

Also the spectrum of the infra-burst time intervals cannot be fit by a single component model. 
As shown in Figure~\ref{fig:spec_infra}, it can be fitted by the same composite model of the burst spectra ($\chi^2$/dof = $6.82/11$). The parameters are highly degenerate, therefore we fixed the  absorption and the emission line to the values found for the bursts.
We obtained a \brem\ temperature $kT=0.4_{-0.1}^{+0.2}$ keV and a photon index  $\Gamma=0.0\pm0.3$, which follows the same trend as a function of CR shown by the burst spectra. 

Both a blackbody plus power law and a disk-blackbody plus power law can fit  well the continuum ($\chi^2$/dof = $6.92/11$ and $6.79/11$, respectively).
The thermal component now accounts for $\sim\!5$\% of the total luminosity $0.3$--$10$ keV, that in the hypothesis of \brem\ emission, is $(9\pm1)\times10^{34}$ \lum.

\subsubsection{Quiescence}

\begin{figure}[htb]
\centerline{\includegraphics[width=1\columnwidth]{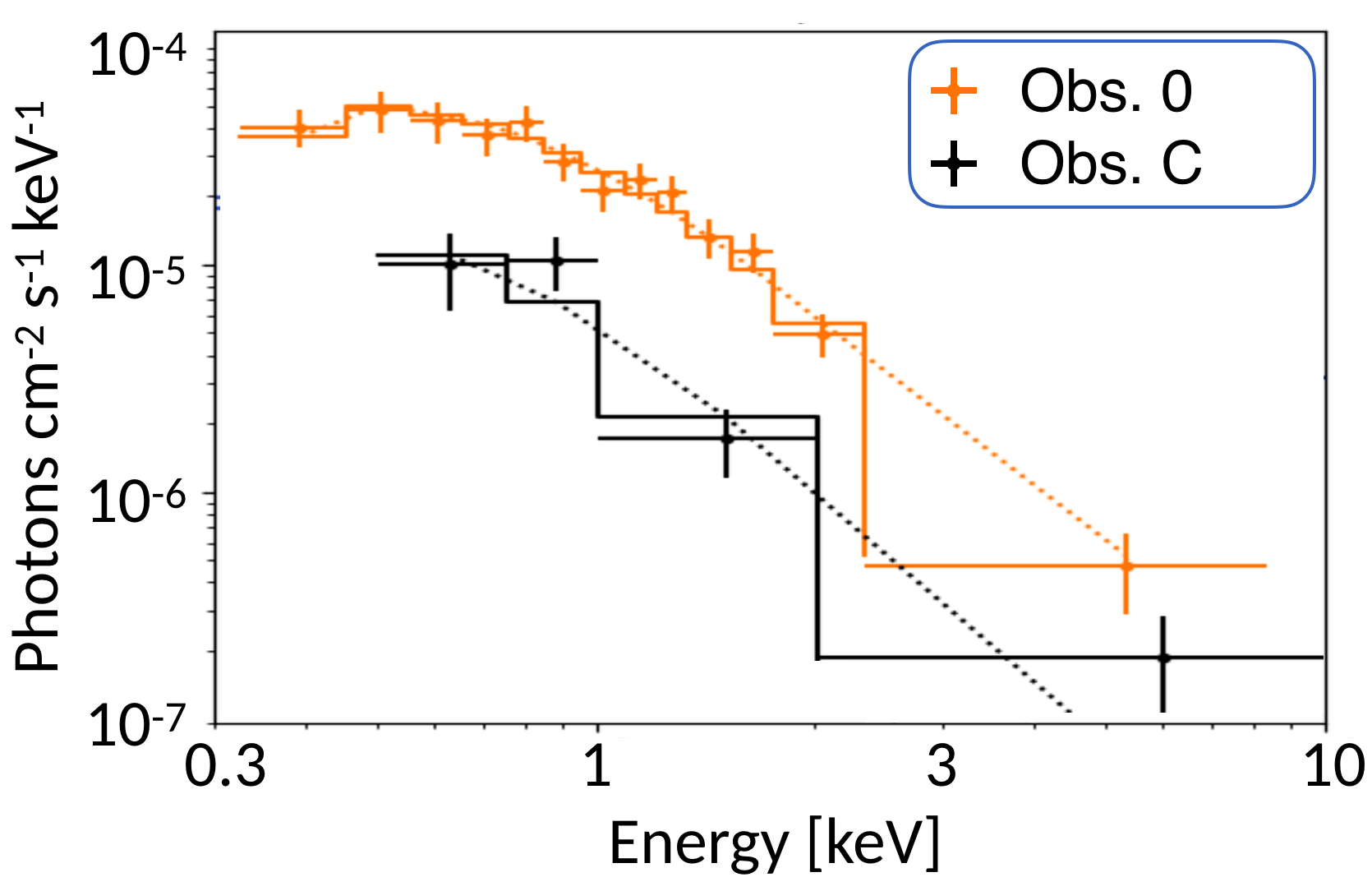}}
\caption{EPIC-pn spectra of obs.\ 0 (orange) and C (black), fitted by a \brem\ model (dotted lines).\label{fig:spec_quiet}}
\end{figure}

Differently from what found for the infra-bursts time intervals, the spectra of the two epochs of quiescence (see Figure~\ref{fig:spec_quiet}) can be fitted by a single thermal component.

The spectrum of Obs.\ 0 is well fitted by either a thermal \brem\ ($N_{\rm H} = (12\pm3)\times10^{20}$ cm$^{-2}$, $kT=1.2\pm0.2$ keV, $\chi^2$/dof = $13.08/17$), or a blackbody ($N_{\rm H} < 50\times10^{20}$ cm$^{-2}$, $kT=0.32\pm0.01$ keV, $R=3.9\pm0.3$ km, $\chi^2$/dof = $11.66/17$). A  power law gives a slightly worse fit ($N_{\rm H} = (27\pm5)\times10^{20}$ cm$^{-2}$, $\Gamma=3.0\pm0.2$, $\chi^2$/dof = $21.30/17$). In the case of \brem\, the luminosity between $0.3$--$10$ keV is $(3.3\pm0.2)\times10^{34}$ \lum.

The spectrum of obs.\ C can be extracted only from the EPIC-pn, and yields only four significant spectral bins. Therefore, we used the best fit thermal \brem\ parameters of obs.\ 0, and varying only the overall normalization, we obtained a total luminosity of $(6\pm2)\times10^{33}$ \lum.

\section{Discussion}\label{sec:disc}

The flaring variability detected in both observations A and B, characterised by flux changes of up to three orders of magnitude on timescales of a few seconds, represents a previously unobserved phenomenon in \src, as well as in other BeXBs. While flaring activity has been documented in a few other high-mass X-ray binaries (HMXBs), it manifests with less extreme properties \citep[see the discussion in][]{2019ApJ...881L..17D}.

\begin{figure*}[htb]
\centerline{\includegraphics[trim=0.cm 0.4cm 1.5cm 1.5cm,clip,width=0.49\textwidth]{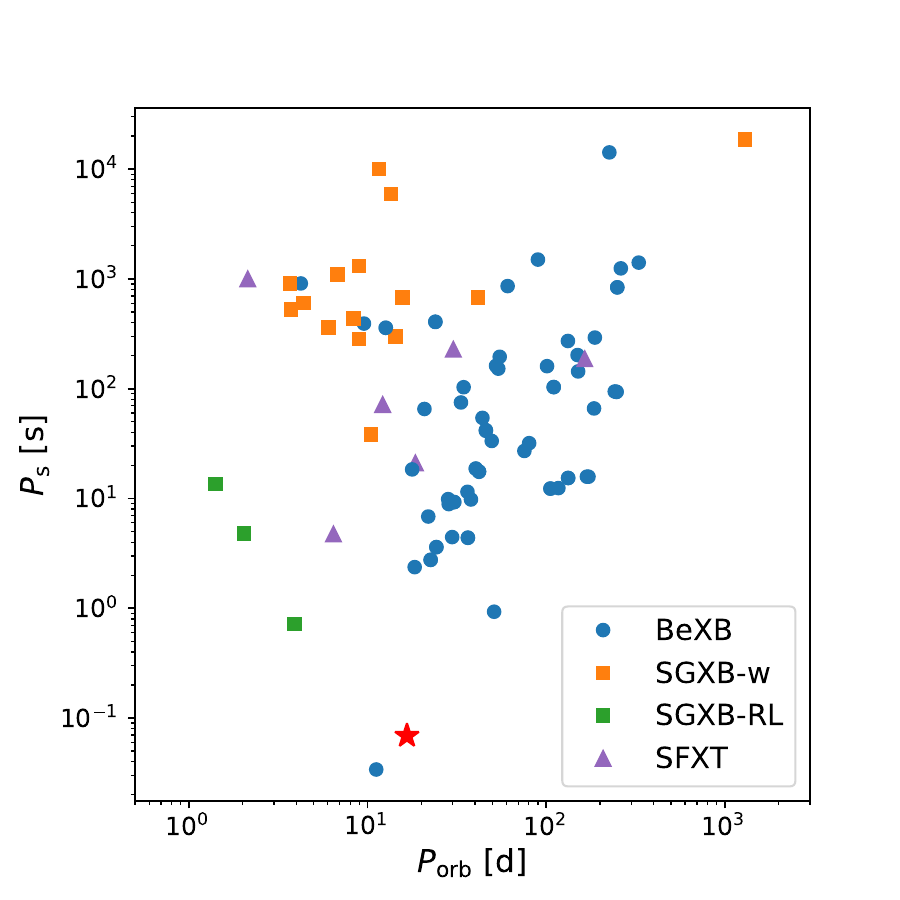} \qquad
\includegraphics[trim=0.cm 0.4cm 1.5cm 1.5cm,clip,width=0.49\textwidth]{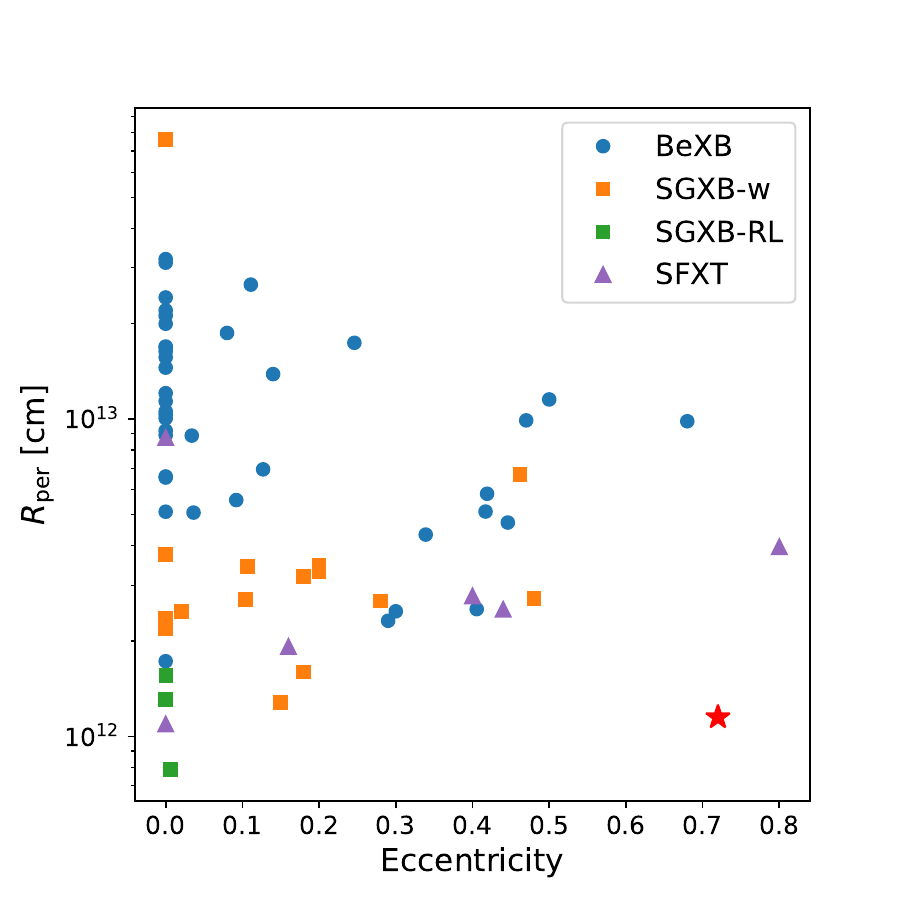}}
\caption{Corbet diagram (left panel) and periastron radius versus eccentricity (left panel) of different types of HMXBs: BeXB (blue dots), Supergiant Fast X-ray Transients (SFXT, violet triangles), Supergiant X-ray binaries whose accretion occurs by stellar wind (SGXB-w, orange squares) or through Roche's lobe overflow (SGXB-RL, green squares). \src\ is indicated with a red star. Data are taken from \citet{2023A&A...671A.149F} and \citet{2011MNRAS.416.1556T}.}\label{fig:bexb}
\end{figure*}

The short and intense bursts in \src\ can be explained in
terms of magnetic gating mechanisms that can occur in accreting sources when the magnetospheric radius (see Equation (2.5) in \citealt{1981MNRAS.196..209D})
\begin{equation}
    R\pdx{M} \simeq 440 \left( \frac{L_X}{10^{38}\mathrm{~erg\,s^{-1}}} \right)^{-2/7} \left( \frac{B}{10^{11}\mathrm{~G}} \right)^{4/7}   \mathrm{~km}
    \label{eq:rm}
\end{equation}
is close to the corotation radius, that for this source is particularly small due to the short spin period of 69 ms:
\begin{equation}
    R\pdx{co} = \left( \frac{GM\pdx{NS} P_s^2}{4\pi^2} \right)^{1/3} \simeq 280 \mathrm{~km}.
    \label{eq:rco}
\end{equation}
Figure~\ref{fig:bexb}, left panel, shows the so-called Corbet diagram for binary systems: among the BeXBs and the HMXBs at large, \src\ is by far the fastest spinning NS along with SAX J0635.2+0533\footnote{SAX J0635.2+053 has $P_s = 33.8$ ms and yet displays anomalously low luminosities (less than $10^{32}$ \lum) during both the active phase and the quiescence. Consequently, due to the high spin-down term ($\dot{P} > 3.8 \times 10^{-13}$ s\,s$^{-1}$), a rotation-powered (rather than an accretion-powered) interpretation is preferred. } \citep{2017A&A...602A.114L}.

A second factor that can contribute to the long-term variability of this system is that its orbit is particularly eccentric ($e=0.72$) in the HMXB framework (Figure~\ref{fig:bexb}, right panel). Due to this, the NS makes a close-by periastron passage and therefore it strongly interacts with the Be wind and its circumstellar disk, that is warped, tilted, and not always present \citep{2017MNRAS.464.4133R,2023MNRAS.523L..75M,2022A&A...661A..22D}.

The spectrum of \src\ during the bursting activity is composed by a broad emission line at 6.4 keV, that can be interpreted as the K$\alpha$  fluorescence line of Fe, plus a power law and a thermal component that is fit by a \brem, a blackbody of radius $R\pdx{BB}\approx 300$ km, or a disk-blackbody. This component contributes $\sim\!30-40$\% to the total flux, an unicum among BeXBs since the intensity of the soft excess for large $R\pdx{BB}$ is usually $5$\% \citep{2004ApJ...614..881H,2012A&A...539A..82L}.

The total X-ray luminosity during the bursts is $L_X \sim 10^{37}-10^{38}$ \lum, well within the typical values expected during a direct accretion regime. It holds as long as $R\pdx{NS} < R\pdx{M} < R\pdx{co}$: according to Eq.~\ref{eq:rm}, this implies that the magnetic field of the NS should be in the range $10^{8} \mathrm{\,G} < B < 10^{11} \mathrm{\,G}$. 

We explored two possibilities to explain the main components of the spectrum in this regime: \textit{ (i)} the hard is the primary component and it is due to an accretion column on the NS, while the soft is a reprocessing component; \textit{(ii)} the soft component is the primary one, due to accreting hot matter, while the hard component is due to Comptonization processes. In both scenarios the expected relative contributions of the two components depend on the details of the assumed model and geometry.

In the infra-bursts phase, the spectrum holds the same features but the luminosity is much lower, $L_X \sim 10^{35}$ \lum. Unless we invoke an unrealistic magnetic field for a BeXB of $10^{9}$ G, the centrifugal barrier should prevent matter to penetrate the magnetosphere, giving the so-called propeller effect. Magneto-hydrodynamic simulations \citep{2005ApJ...635L.165R} suggest that this is true only when the magnetospheric radius is much larger than $R\pdx{co}$ (strong propeller regime), otherwise matter can still (at least in part) penetrate the magnetosphere and accrete onto the NS (weak propeller regime). 

During obs.\ 0 and C we observed a completely constant emission, a purely thermal spectrum, and a weak luminosity of $L_X \sim 10^{33}-10^{34}$ \lum: the system is indeed in a strong propeller regime, where accretion is not possible and the emitted radiation is due to shocked material halted at $R\pdx{M}$.

\section{Summary and Conclusions}\label{sec:conc}

In this contribution, we presented a reanalysis of all the \xmm\ observations of the BeXB \src. In particular, we performed a temporal characterization of these unique short and bright bursts, and we performed a  CR-resolved spectral analysis that showed at least three different emission states.

According to our interpretation, the peculiar behavior of \src\ in the X-ray domain is mainly due to switches between accretion to weak and strong propeller states. 
When $R\pdx{M}$ approaches $R\pdx{co}$, accretion is partly inhibited by the fast rotation of the NS and material 
tends to be ejected from the system instead of being accreted.

During observations A and B, the remarkable variability is due to a hiccup accretion that depends on the local properties of the wind (i.e. accretion rate, velocity and density). The interpretation of the bulk emission during accretion is still open, since no pulsations were detected in this phase (PF<15\%, \citealt{2019ApJ...881L..17D}).
During observations 0 and C, on the other hand, the magnetospheric barrier is closed and we observe shocked material at~$R\pdx{M}$.

\src\ is a remarkable object in the framework of BeXBs and of HMXBs at large; it shows more similarities with other peculiar objects, such as the accreting millisecond pulsar IGR\,J18245--2452 \citep{2014A&A...567A..77F}, the low-mass X-ray binary IGR\,J17407--2808 \citep{2023A&A...674A.100D} (that however have a lower X-ray luminosity), and the Bursting Pulsar GRO\,1744--28 \citep{2018MNRAS.481.2273C} (that exhibits a much lower dynamical range of luminosity).

\section*{Acknowledgments}

MR and SM acknowledge INAF support through the Large Grant ``Magnetars'' (P.I. S. Mereghetti) of the  ``Bando per il Finanziamento della Ricerca  Fondamentale 2022''.










\bibliography{biblio}%

\end{document}